\documentclass[12pt,fleqn]{article}    
\usepackage{umlaute}
\usepackage{epsfig}
\title{
Systematics of the Exclusive Meson Production in the Proton-Proton System in Relativistic Quark-Models
\thanks{Supported in part by the Kernforschungszentrum KFZ J\"ulich \newline
\hspace*{0.6cm} email: mdillig@theorie3.physik.uni-erlangen.de }}

\author{
M. Dillig \\
Institute for Theoretical Physics III  \\
University Erlangen-N\"urnberg, Erlangen, Germany }

\begin{document}
\baselineskip 18pt  
\begin{titlepage}
\maketitle

\begin{abstract}
We investigate the exclusive production of the pseudoscalar mesons $ \pi ^{0}, \eta, \eta^{\prime}, K^{+} $ and of the vector  mesons $ \omega, \phi $ in a microscopic gluon-exchange or instanton model. We describe the baryons as covariant quark - scalar diquark systems with harmonic confinement, thus taking into account center-of-mass corrections and Lorentz contraction in different frames. The excitation of intermediate baryon resonances is accounted by colorless 2-gluon (soft Pomeron) exchange. We find that our model accounts for the systematics of the high precision data on exclusive meson production from various modern proton factories.
\end{abstract}


\end{titlepage}
\newpage
\setcounter{page}{2}
The advent of modern particle accelerators has introduced a new and unprecedented quality of data for electromagnetic and hadronic reactions on the  nanobarn scale: they allow to investigate in detail scattering processes at very large momentum transfers with very low cross sections (1,2). 
\vskip 0.2cm
One of these processes under intensive investigation is the exclusive production of pseudoscalar and vector mesons in proton - proton collisions $ pp \to p B \lambda $
(near the corresponding meson thresholds). As a characteristic feature, all these processes already at threshold involve a very large momentum transfer 
\begin{displaymath}
Q = \frac{1}{2} \; \sqrt{(M_{B} + M + m )^{2} - 4 M^{2} }
\end{displaymath}
(with M and m being the nucleon and the meson mass, respectively. $ M_{B} $ is the mass of the baryon in the final state): the corresponding momentum transfers range from 0.37 GeV/c for $\pi ^{0} $ production up to 1.2 GeV/c for $\phi $ vector meson production. Thus the main goal of these experiments is to test the very short - range nucleon-nucleon dynamics and the structure of mesons and baryons as a step towards an understanding of these processes in the framework of QCD as the theory of strong interaction.
\vskip 0.2cm
For a microscopic formulation of exclusive meson production the choice of the appropriate degrees of freedom is crucial. So far, practically all investigations have been based on meson-exchange models with mesons and baryons as effective degrees of freedom (3). However, in view of the very large momentum transfers mentioned above, which correspond to scales between 0.2 and 0.5 fm in coordinate space, a more appropriate approach presumably has to be based QCD oriented models, i. e. on (constituent) quarks and (nonperturbative) gluons or instantons, as the appropriate degrees of freedom, a field which has not been explored hitherto (except of a single attempt for $ \eta $ production (4)). We believe that the main impact of the high-momentum transfer experiments above is to explore the limit of meson and baryon exchange models and their transition into QCD inspired quark-gluon models.
\vskip 0.2cm
Presently, without a detailed understanding of QCD, in particular of quark-gluon confinement, QCD inspired quark models have to involve a high degree of phenomenology involving various parameters typical for nonperturbative low energy scales (at the range of 1 GeV). Alternatively, however, the advantage of these admittedly crude models is their internal consistency for different meson production processes and, within a (semi) covariant framework, their extension to genuine relativistic features of such reactions, such as the appropriate inclusion of Lorentz boosts. The obvious price one has to pay, is twofold: on the one side economical and feasible models are in general based on the hard - scattering or, equivalently, the Watson-Migdal formalism (5), providing momentum sharing in essentially one very violent collision of the interacting constituents (opposite to production mechanisms dominated by Faddeev - like multiple soft rescatterings (6); in practice, (moderate) corrections in the hard scattering picture are included as soft initial (ISI) and final state (FSI) interactions (7,8). Presently, however, the quality of such a modeling remains questionable: with a complete 3-body Faddeev calculation missing, estimates from coupled channel or perturbative models at least shed some doubt on the adequacy of the WM formalism on a quantitative level (9). Secondly, any present day model building becomes fairly involved, as it has to respect the experience from numerous investigations in meson exchange models. For the route which we follow in this investigation it enforces a formulation of rather detailed, 
but still practical and economical model for the genuine production mechanism.
\vskip 0.2cm
In this note we formulate exclusive meson production in a QCD inspired framework: our degrees of freedom are (constituent) quarks and (nonperturbative) gluons or instantons. As this investigation is a first step towards more detailed and realistic models in future steps, we introduce various simplifications in our model building (they are discussed below), in order to allow for a systematic application to and comparison with data.
\vskip 0.2cm
So far in meson production experiments only total and differential cross sections have been measured as observables (information on spin observables will be obtained in the near future (10)). For the 3-body pB$\lambda$ final state the differential cross section is given in the center-of mass system
\begin{displaymath}
d \sigma = \frac{1}{(2\pi)^{5}} \;  \frac{S}{v_{pp}} \;  \overline{| M_{pp \to pB\lambda}| ^{2}} \;  \frac{d {\bf k}_{p} d{\bf k}_{B} d {\bf k}_{\lambda}}{2 w_{\lambda}} \; \delta (P_{i} - P_{f} )
\end{displaymath}
(with the statistical factor S = 1/2  for the proton-proton final state) with the transition amplitude in the Watson-Migdal parametrization given schematically as 
\begin{displaymath}
M_{pp \to pB\lambda } (K, k, k^{\prime}, k _{\lambda} ) = M(FSI) \ast M(HS) \ast M(ISI)
\end{displaymath}
In our calculation we include initial state interactions via an inelastic reduction factor derived from NN phase shift analysis, while final state interactions are parametrized in a low energy expansion via the scattering length and the effective range; except for the $\eta$-meson production, where the strong $\eta$N interaction is included explicitly, only the proton-proton and the proton-hyperon interaction is included. For details we refer to the literature (3,7,8).
\vskip 0.2cm
The hard scattering production amplitude M(HS) is given as the overlap of the pp initial and pp or pY final state (for K$^{+}$ production) with the corresponding gluon or instanton induced interaction to the mesonic final state
\begin{displaymath}
M(HS) = < p(qQ) Y(qQ) | V(qq \to qq \;( q\overline{q})_{\lambda } )| p(qQ) p (q Q) >
\end{displaymath}
which includes the quark - quark interaction symmetric in the quark indices and proper anti-symmetrization among the quarks and symmetrization among the diquarks, respectively (as discussed below we describe the interacting baryons as quark-diquark $ qQ $ systems).
Right at threshold the partial waves contributing to the transition amplitude are severely restricted: the Pauli-principle, angular momentum and parity conservation restricts the transition from the relative initial pp P-state to the final S-state for the pB system (the pp final state involves only spin singlet, the pY final state both singlet and triplet spin contributions). 
\vskip 0.2cm
We briefly outline the input for the interaction and specify the wave functions of the hadrons.
\vskip 1.0cm
{\large {\bf Interaction }}
\vskip 0.3cm
Schematically, the interaction employed in our model calculation is given in Fig.1. In our calculation we compare two approaches:
\vskip 0.2cm
The nonperturbative instanton induced 6-point $ qq \to qq \; q\overline{q} $ interaction (Fig. 1a) is given for the leading terms in the two-component 
zero-range representation as (11)
\begin{eqnarray*}
 & & V_{int}\; (qq \to qq (q \overline{q}), ijk ) = \\ 
\\
& & \frac{\alpha_{inst}}{m^{4}} \; \lim _{\atop \gamma \to 0 } \; \frac{\partial}{\partial \gamma} \; 
 \left ( \frac{\lambda^{a}\, \lambda^{a}}{4} + \frac{4}{3} \; 
d_{abc} \; \frac{\lambda^{a} \,\lambda^{b}\, \lambda^{c}}{8} \right )\,
\delta \left ( {\bf r}_{i}-{\bf r}_{k} + \gamma \, \mbox{\boldmath$\sigma$}_{k}
\right ) \, \delta \, \left ( {\bf r}_{j} -{\bf r}_{k} \right )
\end{eqnarray*}
(where m denotes the light quark mass, 
$ \lambda $ the standard Gell-Man colour matrices  and $ d_{abc} $ the symmetric SU(3) structure constants). We mention that the structure of the instanton interaction allows both the production of pseudoscaler and vector mesons.
\vskip 0.2cm
Alternatively we start out from the nonperturbative one-gluon exchange in combination with the gluon-induced $ q\overline{q} $ pair creation potential (Fig. 1b), given in the zero-range limit with an effective gluon mass $ m_{g} $ in the two-component representation for the leading terms on the $ q/m $ expansion (12)
\begin{displaymath}
V_{qq} (ij) = \frac{4 \pi \alpha_{s}}{m_{g}^{2}} \cdot \frac{\lambda ^{a} \lambda ^{a}}{4} \; \delta ({\bf r}_{i} - {\bf r}_{j})
\end{displaymath}
and
\begin{displaymath}
V(q \to q\; (q \overline{q}) \; ,ik ) = \frac{4 \pi \alpha _{s}}{m _{g}^{2}} \; 
\frac{1}{m} \; \frac{\lambda^{a} \lambda^{a}}{4} \, .
\end{displaymath}
\newpage
\begin{eqnarray*}
& & \lim _{\gamma \to 0} \quad \frac{\partial}{\partial \gamma} \; ( m^{2} \; (
\delta ({\bf r}_{i} - {\bf r}_{k} + \gamma  (\mbox{\boldmath$\sigma$}_{i} + \mbox{\boldmath$\sigma$}_{k} )  ) + i \; \frac{m_{s}-m}{m_{s}}\;  \delta 
 ({\bf r}_{i} - {\bf r}_{k} + \gamma \mbox{\boldmath$\sigma$}_{k} ) ) \\
\\
&  & + \;  ( 2 \sigma _{k} {\bf p}_{i}\gamma + 2 i \;\frac{1}{a^{2}} \;
 e^{i \mbox{\boldmath$\sigma$}_{k} ({\bf r}_{i} - {\bf r}_{Q} ) \gamma } )\; \delta ({\bf r}_{i} - {\bf r}_{k}) )
\end{eqnarray*}
where we included the spin-dependence in the argument of $\delta$-function in coordinate space, to facilitate the explicit evaluation (above ${\bf r}_{Q} $ results from the gradient on the internal baryon wave function in the initial state; the parameter a reflects the size of the proton; see below). The expression above is extended to the 3-gluon exchange for vector meson production (Fig. 1c).
\vskip 0.2cm
In a further step we account for the coupling to colourless intermediate baryon states (baryon resonances near the meson thresholds seem to dominate $\eta $ and $ K^{+}$ production in the meson-exchange picture (13)) (Fig. 1~d): evaluating the colourless Pomeron exchange as a two-gluon Box diagram (14) we obtain in the zero range limit
\begin{displaymath}
V_{{\bf P}} \,(ij) = \frac{4 \pi \alpha_{{\bf P}}(\Lambda)}{m_{g}^{2}} \; 
\left ( \; \frac{\lambda^{a} \lambda^{a}}{4} \quad \frac{\lambda^{b} \lambda^{b}}{4}  \right )_{0} \cdot \delta ({\bf r}_{i} - {\bf r}_{j} )
\end{displaymath}
where the effective Pomeron coupling constant on the scale parameter $\Lambda$ reflects the size of the constituent quarks in the 2-gluon loop (the index indicates the coupling of the two gluons to color zero). 
\vskip 0.2cm
To obtain some insight in the convergence of irreducible gluon-exchange terms of higher order and their comparison to the instanton interaction, higher order multigluon-exchange contributions were investigated in the heavy quark limit up the 4. order in the running QCD coupling constante $ \alpha_{s}$ (for all technical details and results we refer to work in preparation (15)).
\vskip 0.5cm
{\large {\bf 
Hadron wave functions }}
\vskip 0.2cm
Representing the meson produced as $ q \overline{q} $ states, we compromise in this exploratory study on the structure of the baryons: we represent them as 
quark-diquark objects and keep in practice only the dominant scalar spin-isospin diquark component (16). Then the meson wavefunction as a $ q \overline{q} $ and the baryon as a $ q-(qq)$ object are given schematically in the standard form as
\begin{displaymath}
\phi (ij) = \phi ({\bf r}_{ij}) \; \chi_{spin} \;  \chi_{flavour}  \; \chi_{colour} \; ,
\end{displaymath}
$ \phi $ is four-component relativistic wave function (including the large and small 
component). For an harmonic confining kernel the large component of the boosted wave function in a system moving with momentum P along the z - axis is given for the nucleon - and correspondingly for the meson - in its ground state as (17)
\begin{displaymath}
\phi (z, \mbox{\boldmath$\rho$} ) = N \,\exp \left ( - \frac{1}{2a^{2}} \; \left ( \frac{z^{2}}{\lambda^{2}} + \, \mbox{\boldmath$\rho$}^{2} \right ) \right )
\end{displaymath}
where $ \mbox{\boldmath$\rho$}$ denotes the component perpenticular to the z-axis, while a is the size parameter of the hadron; the quantity
$ \lambda = M / \sqrt{{\bf P}^{2} + M^{2}} $ reflects the Lorentz quenching of the z-component of the hadronic wave function for momenta $ {\bf P} = (0,0,P$). Furthermore we represent the s-wave (negative parity)  $1/2^{-}$ baryon resonances in $ \eta $ and $K^{+} $ production as an orbital p-wave excitation of the q - Q system (18).
\vskip 0.2cm
Im presenting a set of characteristic results, we stress again our main goal to reproduce qualitatively the major trends in the various near threshold cross sections $ pp \to p\,B \,\lambda $ with a minimal set of parameters, rather than to produce an optimal fit to the data.  In our survey we cover the energy dependence of the total cross section for the various mesons, where a substantial amount of data is presently established, i. e. $ \pi^{0}, \eta, \eta^{\prime}, K^{+}, \omega $ and $ \phi $ production in the proton - proton system (Fig. 2; the data are taken from refs. 19-24). 
We fix the parameters in our calculation from current values in the literature: the coupling constants 
$\alpha_{s} = 1.7, \alpha_{inst} = 2.0 $ and $\alpha_{{\bf P}} = 2.2 $, the masses $ 
m_{u} = m_{d} = 330 $ MeV,  $m_{Q} = 600 $ MeV,  $m_{g} = 660 $ MeV and $ \Lambda = 450 $ MeV,  together with 
the size parameters $a_{p} =  a_{\Lambda} = 0.6 $ fm  and $a_{\lambda} = 0.5 $ fm. 
Furthermore, for the absolute normalization of the cross section we normalize our prediction to the cross section for $\pi^{0}$ production at the excess energy of Q = 9.6 MeV. Then without further adjustments we find that for the pseudoscalar and the vector mesons (with the inclusion of the colorless Pomeron exchange) the major trends of the date are qualitatively reproduced. In view of the various basic uncertainties of the models, particularly with respect to the absolute normalization of the various cross sections, it is difficult to discriminate quantitatively between the (nonperturbative) gluon exchange and instanton induced meson production (within the accepted range of strength parameters the two parametrizations provide qualitatively similar results for near-threshold), however, for resonance dominated meson production, such as the $\eta$ and the $K^{+}$ meson, the instanton induced interaction underestimates the data by typically a factor 4 (compare Fig. 3(a) below). In performing the comparison with the data Lorentz quenching is crucial for a qualitative agreement with the data: alternatively, the radius parameter for the incoming protons has to be reduced substantially from its value obtained from spectroscopy, where the proton wave function enters in its rest frame.
\vskip 0.2cm
For $ pp \to pp \eta $ the influence of the Pomeron exchange with the excitation of the N*(1535) is demonstrated explicitly (Fig.~3~a). Evidently, the cross section from the pure 2-gluon or instanton exchange is enhanced by typically a factor 4 or more, provided Pomeron exchange is included. Clearly, the calculation is still preliminary, as it involves significant uncertainties in the evaluation of the 2-gluon loop; it evidently does not match the quality of sophisticated meson exchange calculations (13), however, it is found that - opposite to the production of other heavy mesons, which are presumably not resonance dominated,  Pomeron exchange and $N^{*}$ excitation is indispensable for $\eta$ (and $K^{+}$) production. In addition we find that qualitatively the model reproduces the angular distributions for the $\eta $ meson at different excess energies (25)(Fig.~3~b).
\vskip 0.2cm
Opposite to the 2-gluon exchange for pseudoscalar mesons, vector-meson production is dominated by the 3-gluon exchange contribution. The major trends for $ \omega $ and $ \phi $ production are reproduced qualitatively, though the microscopic structure of the production processes are rather different: for the $ \phi $ meson as a pure $ \overline{ss}$ state only the direct production via the 3-gluon exchange contributes (for the $ \omega $ production the mechanism 
involves additional contributions). Clearly, further investigations are here necessary and interesting, in particular with respect of the $ \phi /\omega $ ratio in comparison with the prediction of the OZI rule (26) and with the respect to the strange content of the proton (presently details are worked out and presented in a separate publication). 
\vskip 0.2cm
We summarize the brief survey of our model calculations. In a simple gluon and instanton exchange model, together with a relativistic quark-diquark model for the baryons involved in the production process, we qualitatively account for the major trends of the data. Though the model predictions presently do not match the successes of sophisticated meson exchange calculations for the various meson production processes, their main advantage is the consistency of the results with a very small set of adjustable parameters. The success of the model is not surprising: by construction it simulates the "nucleonic" and the "mesonic" currents in the meson-exchange models and allows, if extended by colorless (correlated) multi-gluon (Pomeron) exchanges, explicitly  the excitation of colorless baryon resonances, such bridging the gap to resonance dominated production mechanisms. 
\vskip 0.2cm
Clearly, for a quantitative comparison and extraction of parameters the model approach is by far too preliminary. Quite evident improvements are the inclusion of finite range effects in the interaction, an inclusion of vector - diquarks in the baryon wave functions (which is inevitable for associated $ \Sigma K^{+}$  production) or a more realistic treatment of the final state interactions, particularly in the meson-baryon system, to name only obvious extensions. Beyond that only a systematic application of the model to all existing observables  and, to explore its predictive power, to more sensitive observables, like the analyzing power or spin transfer coefficients (where data are expected in the near future), will ultimately  contribute to bridge the gap between effective mesonic and baryonic degrees of freedom and QCD as the adequate framework of strong interactions.

\newpage

\begin{figure}
\centerline{
\epsfig{file=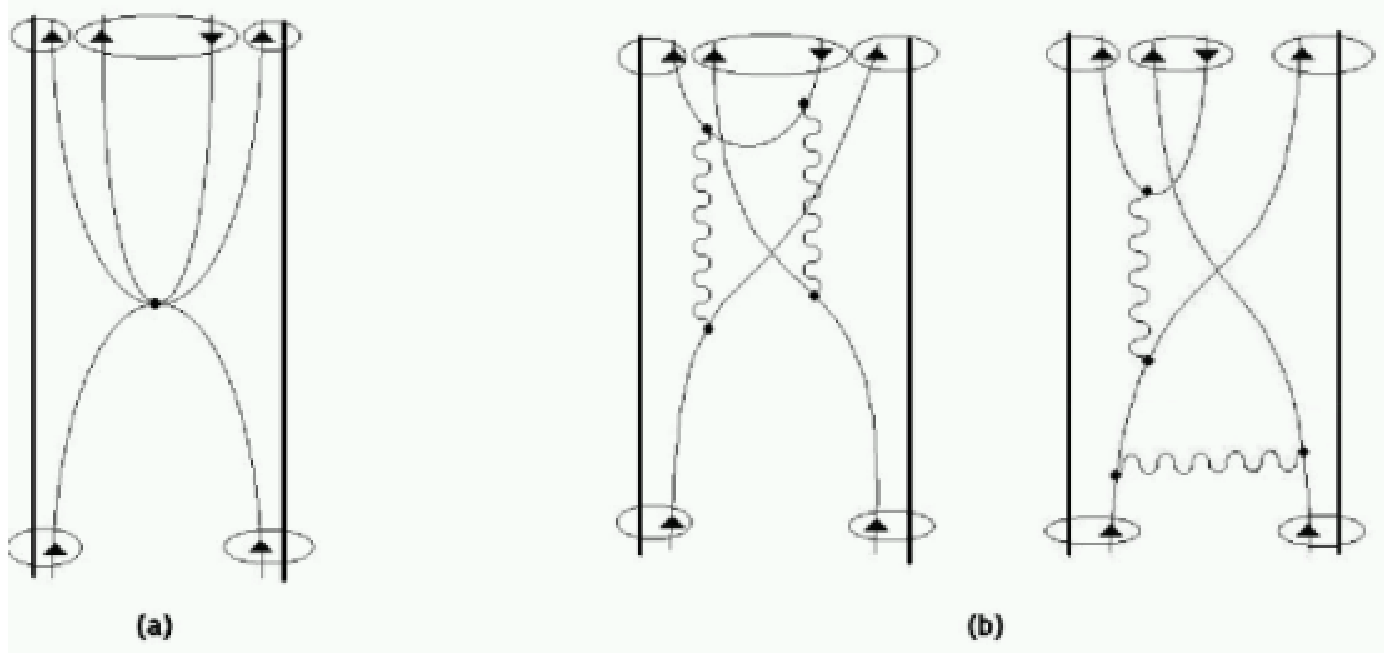}}  
\centerline{
\epsfig{file=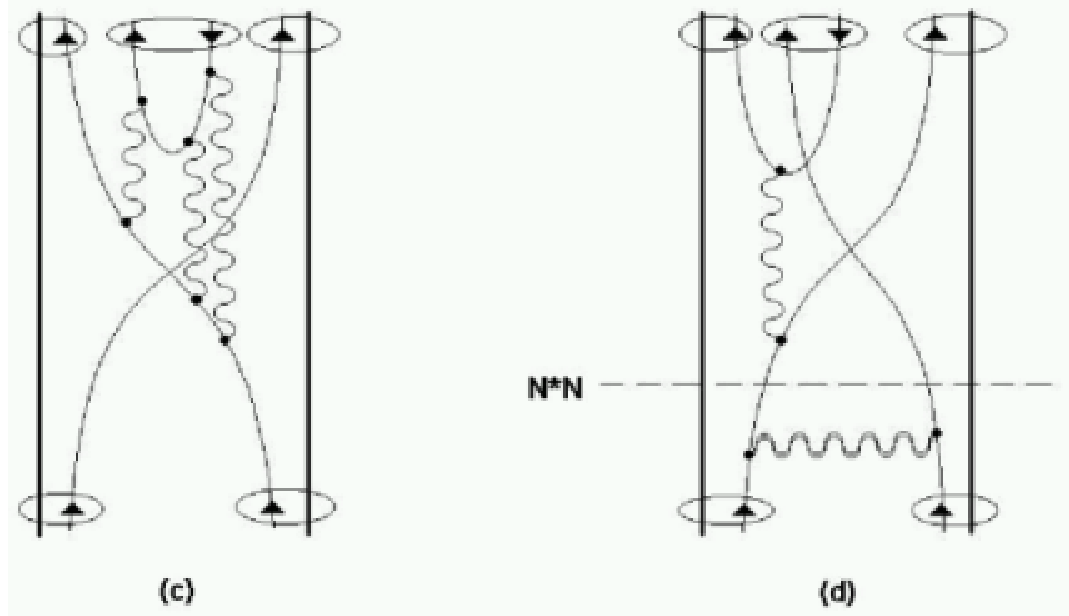}} 
\caption{Typical diagrams for the $qq \to qq (q\overline{q})$
production operator: \newline  
(a) instanton induced 6-quark interaction; \newline
(b) two-gluon exchange and rescattering mechanism;     \newline
(c) 3-gluon exchange mechanism for vector meson production;   \newline
(d) correlated, colorless two-gluon (Pomeron) exchange (the dashed
line indicates the excitation of an intermediate $N^{*}N $ system)}  
\end{figure}

\stepcounter{figure}

\begin{figure}

\vspace*{-1cm}
\parbox{0.49\linewidth}{\hskip -.4\linewidth
\epsfig{file=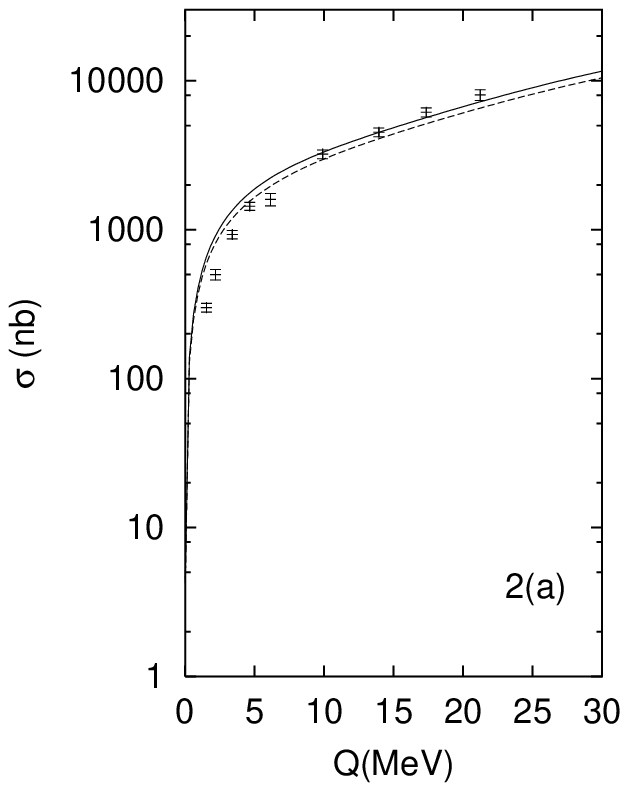,width=2\linewidth,height=\linewidth}}
\hfill
\parbox{0.49\linewidth}{\hskip -.3\linewidth 
\epsfig{file=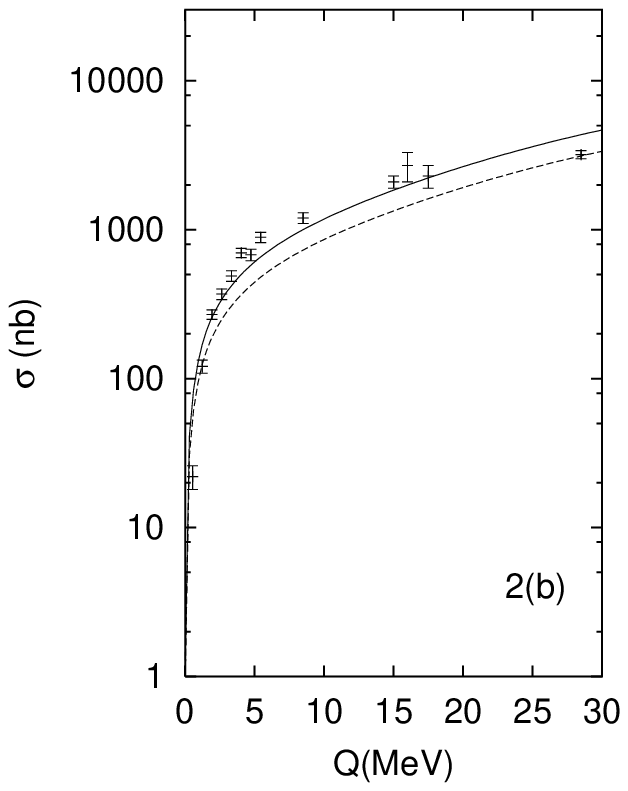,width=2\linewidth,height=\linewidth}}

\parbox{0.49\linewidth}{\hskip -.4\linewidth
\epsfig{file=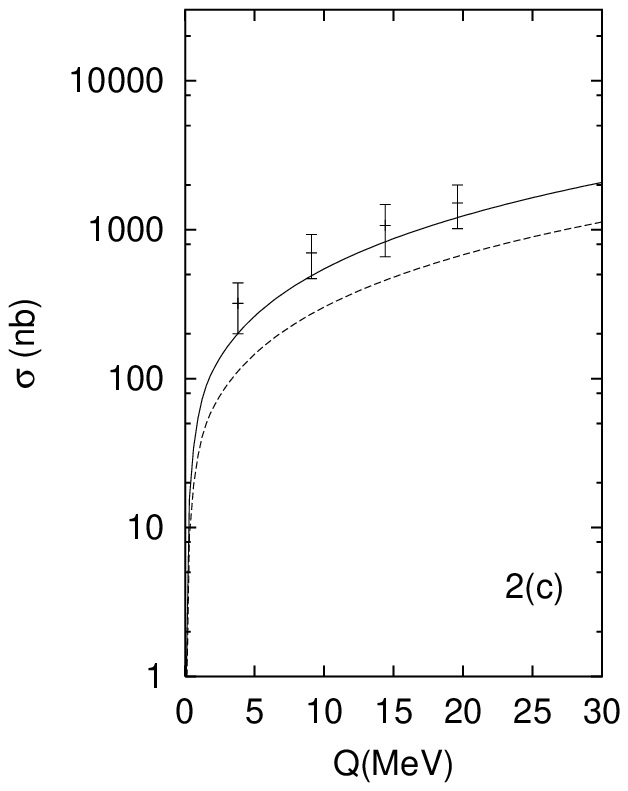,width=2\linewidth,height=\linewidth}}
\hfill
\parbox{0.49\linewidth}{\hskip -.3\linewidth 
\epsfig{file=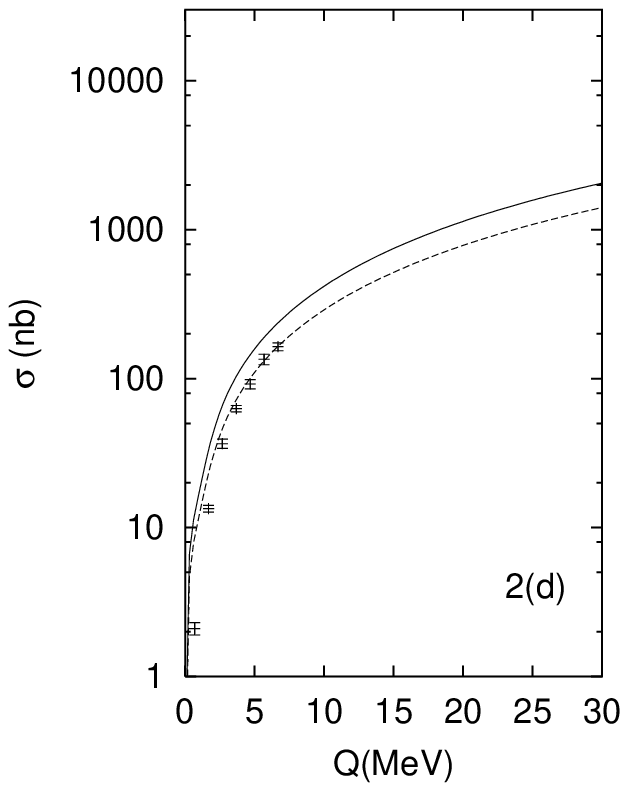,width=2\linewidth,height=\linewidth}}

\parbox{0.49\linewidth}{\hskip -.4\linewidth
\epsfig{file=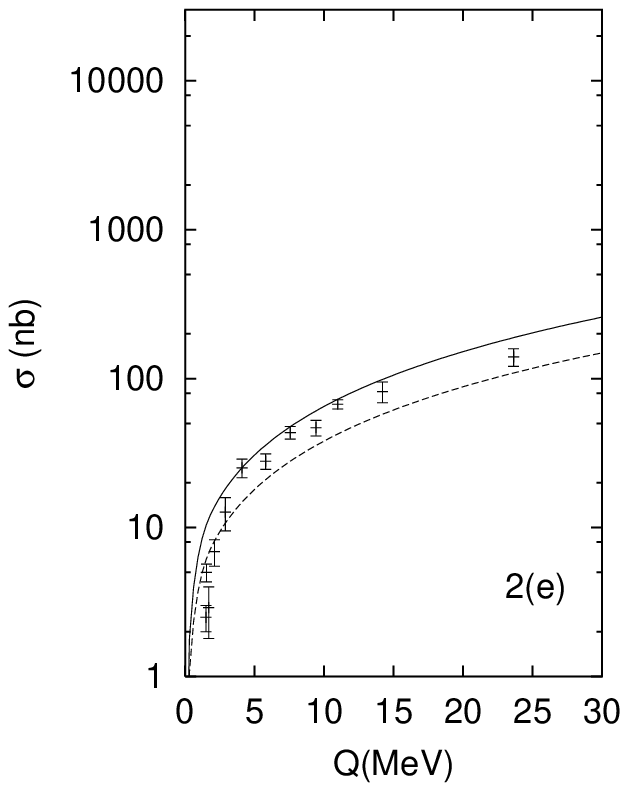,width=2\linewidth,height=\linewidth}}
\hfill
\parbox{0.49\linewidth}{\hskip -.3\linewidth 
\epsfig{file=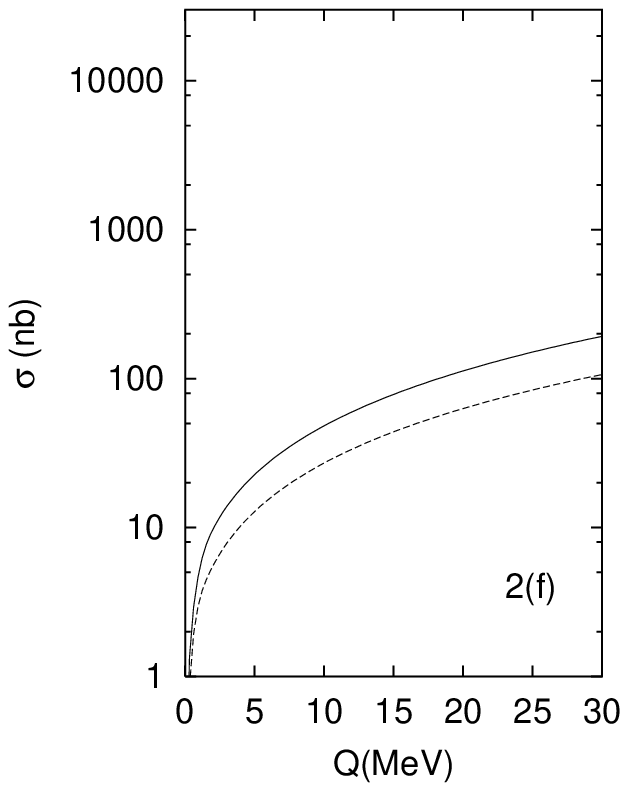,width=2\linewidth,height=\linewidth}}

\caption{Energy dependence of the total $pp \to pB \lambda $ cross
section as a function of the excess energy $Q $. The full and dashed
lines denote the complete gluon exchange contribution for $a \equiv
0.60$ and $0.65 fm$, respectively. The data are taken from ref. (19-24)}
\end{figure}

\begin{figure}

\parbox{0.49\linewidth}{\hskip -.5\linewidth
\epsfig{file=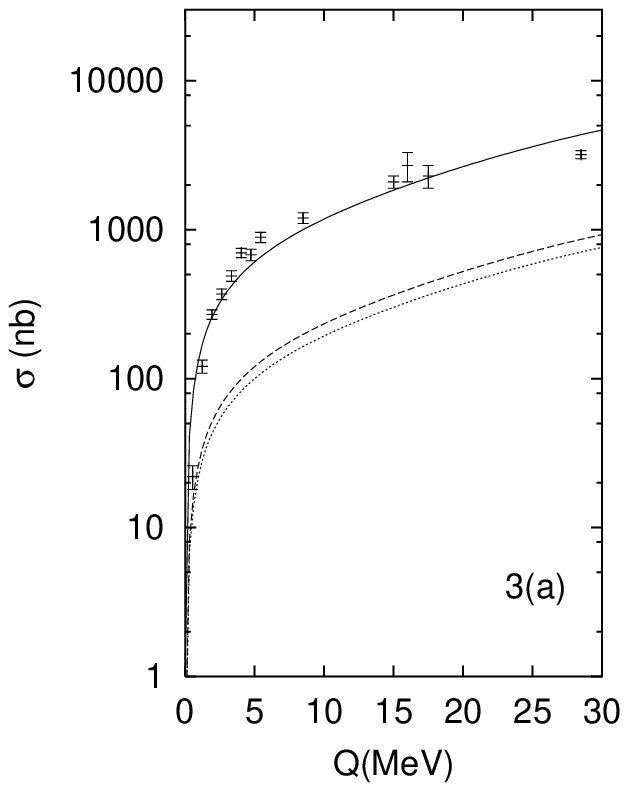,width=2\linewidth,height=\linewidth}}
\hfill
\parbox{0.49\linewidth}{\hskip -.3\linewidth
\epsfig{file=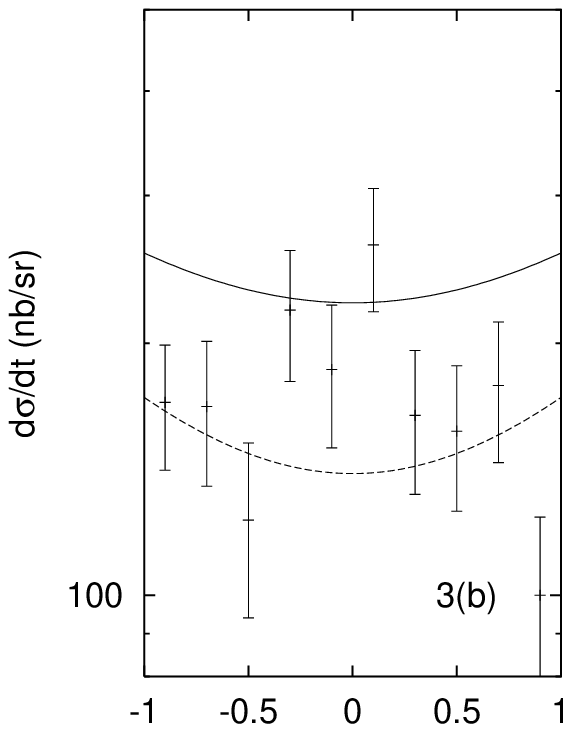,width=2\linewidth,height=\linewidth}}

\parbox{0.49\linewidth}{\hskip -.5\linewidth
\epsfig{file=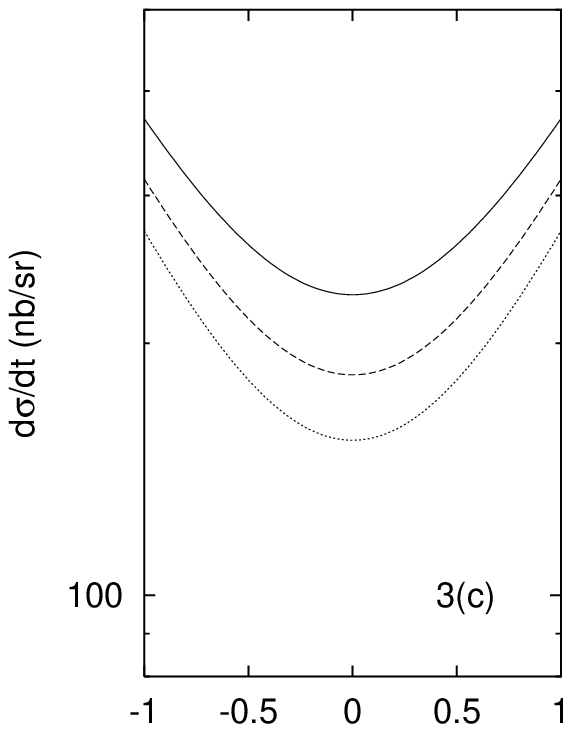,width=2\linewidth,height=\linewidth}}
\hskip .3cm
\parbox{0.5\linewidth}{
\caption{(a) Comparison of the complete gluon exchange interaction
(including Pomeron exchange, comp.  
Fig.~1(b,d)) (full  line) with the 2-gluon exchange and the instanton induced
 contribution (dashed and dashed-dotted line, respectively); 
\newline
(b) for the proton final state in the reaction $pp \to pp \eta $ at  Q
= 16  MeV  for a = 0.60 and 0.65 fm and m$_{Q}$ = 600 MeV (full and
dashed line);  
\newline
(c) angular p distribution at Q = 37 MeV, for a = 0.65 fm and m$_{Q}$
= 575, 600 and 625 MeV (full, dashed and dashed-dotted line,
respectively). Compared are data for the excess energy 16 MeV from
ref. 25.}}  

\end{figure}

\end{document}